\documentclass[preprint,12pt]{elsarticle}

\usepackage{amssymb}

\usepackage{lineno}

\usepackage{lmodern}
\usepackage[hidelinks]{hyperref}

\usepackage{todonotes}

\newcommand{\mod}[1]{#1}

\usepackage{listings}
\usepackage{color,url}

\definecolor{codegreen}{rgb}{0,0.6,0}
\definecolor{codegray}{rgb}{0.5,0.5,0.5}
\definecolor{codepurple}{rgb}{0.58,0,0.82}
\definecolor{backcolour}{rgb}{0.95,0.95,0.92}

\lstdefinestyle{mystyle}{
    backgroundcolor=\color{backcolour},   
    commentstyle=\color{codegreen},
    keywordstyle=\color{magenta},
    numberstyle=\tiny\color{codegray},
    stringstyle=\color{codepurple},
    basicstyle=\ttfamily\footnotesize,
    breakatwhitespace=false,         
    breaklines=true,                 
    captionpos=t,                    
    keepspaces=true,                 
    numbers=left,                    
    numbersep=5pt,                  
    showspaces=false,                
    showstringspaces=false,
    showtabs=false,                  
    tabsize=2,
    mathescape=false
}

\usepackage{tikz}
\usetikzlibrary{shapes.geometric, arrows, calc}

\tikzstyle{startstop} = [rectangle, rounded corners, 
minimum width=3cm, 
minimum height=1cm,
text centered, 
draw=black, 
fill=red!30]

\tikzstyle{io} = [trapezium, 
trapezium stretches=true, 
trapezium left angle=70, 
trapezium right angle=110, 
minimum width=3cm, 
minimum height=1cm, 
text centered, 
text width=5cm, 
draw=black, fill=blue!30]

\tikzstyle{process} = [rectangle, 
minimum width=3cm, 
minimum height=1cm, 
text centered, 
text width=5cm, 
draw=black, 
fill=orange!30]

\tikzstyle{decision} = [diamond, 
minimum width=3cm, 
minimum height=1cm, 
text centered, 
draw=black, 
fill=green!30]
\tikzstyle{arrow} = [thick,->,>=stealth]

\usepackage{multirow}
\usepackage{multicol}

\usepackage{soul}

\newcommand*{\etal}{\textit{et al.}}

\journal{Information and Software Technology}

\begin{document}

\begin{frontmatter}



\title{Automatic Build Repair for Test Cases using Incompatible Java Versions\tnoteref{copyright}}
\tnotetext[copyright]{\copyright\:2024. This manuscript version is made available under the CC-BY-NC-ND 4.0 license 
\url{https://creativecommons.org/licenses/by-nc-nd/4.0/}.}


\author[add]{Ching Hang MAK}
\ead{chmakac@connect.ust.hk}
\author[add]{Shing-Chi CHEUNG}
\ead{scc@cse.ust.hk}

\affiliation[add]{
    organization={Department of Computer Science and Engineering, The Hong Kong University of Science and Technology}, 
    city={Hong Kong},
    country={China}
}

\begin{abstract}
\paragraph{Context} Bug bisection is a common technique used to identify a revision that introduces a bug or indirectly 
fixes a bug, and often involves executing multiple revisions of a project to determine whether the bug is present 
within 
the revision. However, many legacy revisions often cannot be successfully compiled due to changes in the programming 
language or tools used in the compilation process, adding complexity and preventing automation in the bisection process.

\paragraph{Objective} In this paper, we introduce an approach to repair test cases of Java projects by performing 
dependency minimization. Our approach aims to remove classes and methods that are not required for the execution of one 
or more test cases. Unlike existing state-of-the-art techniques, our approach performs minimization at source-level, 
which allows compile-time errors to be fixed.

\paragraph{Method} A standalone Java tool implementing our technique was developed, and we evaluated our technique 
using subjects from Defects4J retargeted against Java 8 and 17.

\paragraph{Results} Our evaluation showed that a majority of subjects can be repaired solely by performing minimization,
including replicating the test results of the original version. Furthermore, our technique is also shown to achieve 
accurate minimized results, while only adding a small overhead to the bisection process.

\paragraph{Conclusion} Our proposed technique is shown to be effective for repairing build failures with minimal 
overhead, making it suitable for use in automated bug bisection. Our tool can also be adapted for use cases such as bug 
corpus creation and refactoring.

\end{abstract}




\end{frontmatter}


\section{Introduction}
\label{intro}

Bug bisection is a common technique used by software developers to identify the commit which introduced a regression in 
a software project, and was first outlined to minimize the effort required in the process of identifying and fixing a 
regression \cite{ness1997regression}. Since manually performing bisection can be a repetitive process, many version 
control systems that implement bisection capabilities also implement automatic bug bisection, such as Git 
\footnote{\url{https://git-scm.com/}}. Generally speaking, automated bug bisection works by using a developer-provided 
command or script to automatically build and/or execute tests, and use the result of the execution to determine whether 
the snapshot manifests the bug or not, thus requiring no manual intervention during the bisection process. 

Because automated bug bisection relies on a script to compile and execute any relevant test cases, the effectiveness of 
this technique is decreased when bisection reaches a snapshot that cannot be successfully compiled, since the snapshot 
will either need to be skipped or, if the version control system does not support skipping revisions, the bisection will 
need to be performed manually. This issue is further amplified when a project's history uses different versions of a 
programming language over its history, leading to the older snapshots of the project being uncompilable because of 
language features or library APIs being removed over time.

\begin{lstlisting}[numbers=none,label={lst:compile-failed-err-msg},caption={An snippet of the compilation failure 
message due to standard library changes.}]
    [javac] src/test/java/com/fasterxml/jackson/dataformat/xml/failing/Issue37AdapterTest.java:7: error: package javax.xml.bind.annotation does not exist
    [javac] import javax.xml.bind.annotation.*;
    [javac] ^
    [javac] src/test/java/com/fasterxml/jackson/dataformat/xml/failing/Issue37AdapterTest.java:8: error: package javax.xml.bind.annotation.adapters does not exist
    [javac] import javax.xml.bind.annotation.adapters.*;
    [javac] ^
\end{lstlisting}

Listing \ref{lst:compile-failed-err-msg} demonstrates a compilation error caused by building a legacy project snapshot 
using Java Development Kit (JDK) version 11. The root cause of this issue is the removal of Java API for XML Processing 
(JAXP) in Java 11, leading to the compilation error related to missing packages and declarations. This also shows that 
all project snapshots utilizing JAXP cannot be compiled using JDK versions 11 or above, rendering automated bisection 
ineffective.

In this paper, we present \textit{test dependency minimization}, a technique used for automatically repairing 
compilation-related build failures. Given a broken snapshot and relevant test case(s) as its input, test dependency 
minimization utilizes reachability analysis and whole-project context to automatically minimize the classes, methods, 
and field declarations (hereinafter referred to as \textit{program declarations}) required to compile and execute the 
test cases, removing the source of the compilation error in the process and thus repairing the compilation failure of 
the snapshot. To increase the effectiveness of the repair process, we propose several techniques that increase the 
number of removable declarations while also minimizing the runtime of our technique, allowing this tool to be run as 
part of a bisection script with minimal runtime overhead.

In addition to aiding in automated bisection, test dependency minimization can also be applied to the scenario of bug 
corpus collection, allowing otherwise uncompilable candidate subjects to be included in a dataset and therefore 
diversifying the dataset by the inclusion of more subjects.

We evaluated this technique on 130 and 951 subjects from Defects4J using Java 8 and 17 respectively, and the results 
show that the technique can repair compilation in all instances, with a further 91\% and 84\% of subjects correctly 
replicating the execution result of the relevant test cases respectively. We also demonstrate that test dependency 
minimization takes up to 20 seconds to execute, showing that the increase in runtime is small compared to performing 
manual bisection.

The contributions of this paper can be summarized as follows:

\begin{itemize}
\item We demonstrate the reasons for build failure after upgrading the Java compiler version, and identified 4 
categories of compilation errors caused by compiler upgrade: Changes to the Java Language, changes to the Java standard 
library, unsupported encoding, and unsupported build tools.

\item We propose a technique that, through the removal of unused classes and methods, eliminates the source of 
compilation errors and thus allows the snapshot to be compiled and/or executed. We also propose a reachability model 
that accurately determines whether a program declaration is required in the compilation process. To our knowledge, this 
is the first work that performs compilation error repair via the use of minimization, at the same time preserving 
program behavior.

\item We provide an implementation of the aforementioned technique which implements static analysis-based dependency 
analysis and minimization. The implementation of the technique is open-sourced and can be found on GitHub 
\footnote{\url{https://github.com/Derppening/test-dependency-minimization}}.
\end{itemize}

The rest of the paper will be structured as follows: Section \ref{bg} will describe the background and motivations of 
this work. Section \ref{methodology} will describe a high-level overview and implementation details of the proposed 
technique. Section \ref{eval} will describe the testing methodology and analyze the effectiveness of the technique. 
Section \ref{discussion} will discuss the possible use cases of the technique, as well as outline possible threats to 
validity in the experimentation. Section \ref{related-works} will discuss related works, before concluding the paper in 
Section \ref{conclusion}.

\section{Background}
\label{bg}

\subsection{Motivating Example}

We will utilize Listing \ref{lst:compile-failed-err-msg} to describe the key observations that drive the basis of our 
technique, and outline the technical challenges associated with the technique.

\subsubsection{Potential Workflow and Issues}

To provide additional context to the error, the motivating example is taken from a snapshot in the Jackson XML 
dataformat library \footnote{Git Commit ID 
\href{https://github.com/FasterXML/jackson-dataformat-xml/commit/81f38e1985bdfafdbe02e32dfb5ccb200fc64eae}{\nolinkurl{
81f38e1}}}, and is aimed to fix a bug designated \texttt{\#180} 
\footnote{\url{https://github.com/FasterXML/jackson-dataformat-xml/issues/180}} in the GitHub repository. Along with 
the fix, this commit also introduces changes to four test cases, each of them adding assertions to verify that 
additional constraints related to the reported bug are met.

Let us assume that a developer would like to perform a bisection of the source of this issue. If the development 
environment is set to use Java 11 or above, the compilation stage of the bisection script will fail due to ``package or 
class can not be found'' errors as previously described. In this case, the only solution is for the developer to 
install JDK 8 in their development environment, and force the build system used by the snapshot to use this version of 
Java throughout the bisection process.

\mod{However, this solution is no longer viable with projects upgraded to Java 9 or above, which introduced a new 
policy where only the most recent 3 Long-Term Support versions of Java will be supported, as well as an accelerated API 
deprecation-removal cycle. This policy means that for libraries such as Mockito 
\footnote{\url{https://github.com/mockito/mockito}}, with project snapshots ranging from Java 5 to 11, bisection of a 
newly-discovered fault will require at least JDK versions 1.8 (which supports compiling down to Java 5) and 11 (which 
supports compiling down to Java 6).}

\mod{Under these limitations}, a developer may then choose to workaround this problem by implementing a script that 
determines the required Java version for a snapshot. However, this introduces a new set of problems.

\begin{itemize}
\item \textbf{Determination of Target JDK:} \mod{While most build systems allow developers to specify which JDK version 
the project must be compiled with, this declaration is optional and will default to the system's default Java version 
for compilation. These projects will therefore require a tool that can infer the required Java version for a project 
snapshot, and this tool will need to be updated whenever a new Java version is released.}

\item \textbf{Emulating Legacy JDKs:} \mod{While older versions of JDK are still available for download, these versions 
do not support installation on modern operating systems. Moreover, working around this issue by using newer JDKs at a 
lower source level can still cause compilation errors as older versions of the Java Standard Library is not bundled 
with newer JDKs, potentially causing resolution ambiguities with new APIs.}
\end{itemize}

\subsubsection{Observations of Interest}
\label{observation-of-interest}

To address the problem described in the previous subsection, we can utilize two key observations from the snapshot and 
the nature of the build failure to formulate a solution.

\begin{enumerate}
\item \textbf{Usage of Test Suite in Bisection:} When performing bisection on a bug, it is often common practice to 
use a subset of the project test suite to check for the existence of the bug in a specific snapshot 
\cite{ness1997regression}.

\item \textbf{Scope of Unit Tests:} Many software projects utilize unit tests to verify the correctness of different
parts of the project. Unit tests are often preferred because these tests only target a unit of a program, meaning that 
the source of a test failure can be quickly localized based on the target unit of the failing test 
\cite{xie2007towards}.

With the motivating example, the JAXB library provides annotations for databinding as well as classes acting as 
the ground truth for Jackson XML's test cases. However, none of the modified test cases utilize the JAXB library, 
therefore test cases that depend on the JAXB library can be safely removed to allow our test case to be executed.
\end{enumerate}

Based on these observations, we can make the inference that since a subset of the test suite will be used during the 
bisection process for determining the presence of the bug, and since each unit test only exercises a small subset of 
the project, it is possible to repair the compilation error by removing all parts of the project snapshot that are not 
used by any test case in the bisection process. This technique will be referred to as \textit{minimization} throughout 
this paper and forms the basis for our proposed technique.

\subsection{Technical Challenges}
\label{technical-challenges}

We note the following challenges when designing and implementing a technique based on the removal of unused program 
declarations for repairing compilation failures.

\begin{itemize}
\item \textbf{Accurately Determining Necessary Declarations}

Due to the structured nature of source code, more program declarations need to be retained to allow it to be compiled 
successfully. At the same time, since the goal of test dependency minimization is to repair broken snapshots in newer 
versions of Java, the technique must also be designed to minimize the set of retained declarations to eliminate the 
cause of the compilation error. This means that there will be an optimal minimization result that minimizes the set of 
retained declarations while still being compilable. Since the inclusion of any unneeded declaration may cause the 
source of the compilation error to be retained, while the exclusion of any needed declaration may introduce new 
compilation or runtime errors, the proposed technique should reach as close to this result as possible.

To address this challenge, we note that performing minimization under a coarser granularity includes more redundant 
declarations, as we will be unable to make any inferences on the reachability of its class members, and thus we must 
treat all members within the class as reachable. Therefore, we chose to perform minimization on member granularity, as 
this is the most granular level where reachability for each declaration can be accurately inferred without the use of 
any runtime information.

We also note that the full context of the project is available as the input for our technique, meaning that we can 
exploit this to improve the accuracy of minimization. Therefore, we devise a two-phase approach for the minimization 
process, by first marking all declarations possibly needed in the compilation and/or execution of the test case, 
followed by running reachability analysis using the global context to accurately decide whether a declaration is 
required by the reachability of its dependent declarations. This process is further discussed in Section 
\ref{minimization-phases}.

\item \textbf{Devirtualizing Virtual Method Calls} 

One of the major difficulties in accurately minimizing a Java program is determining the set of methods that may be 
invoked. Since Java uses dynamic method lookup for all non-static method calls by default, a method call may invoke the 
static target of the method call or an overriding method in one of its subclasses. This is especially challenging when 
the static type of the method scope is a library type, as a naive approach will cause many overriding methods to be 
included, increasing the number of redundant declarations in the minimization result.

To address this challenge, we exploit the fact that the full context of the project is available, meaning that the exact 
types of the method scope can be narrowed down. We also note that some types of expressions can have a more constrained 
type than what is inferred by the Java type system, which further reduces the set of candidate methods during dynamic 
dispatch. This optimization is further discussed in Section \ref{opt:mindispatchset}.
    
\item \textbf{Preserving Diagnosibility} 

Since the proposed technique is based on static analysis, the technique will suffer from the same drawbacks as other 
static analysis techniques, such as being unable to accurately process classes and methods used via the Java Reflection 
APIs, and therefore will require manual intervention. Under those circumstances, our technique should be able to output 
information to help developers identify the cause of the issue and perform manual fixes.

To address this challenge, we provide a mode of operation for users where unreachable program declarations will have 
their body replaced with an \texttt{AssertionError} with the context of the unreachable method. This option is further 
discussed in Section \ref{sweep-phase}.
\end{itemize}

\paragraph{Key Idea} During a bisection process, the bisection may reach a snapshot that causes a build failure when 
executing under an unsupported JDK version. As discussed in \ref{observation-of-interest}, since only a small subset of 
test cases and their dependent program declarations are necessary to determine the correctness of the snapshot, 
declarations that are not used in the compilation or execution of the test cases can be removed. This will likely lead 
to the removal of the source of the compilation error, allowing the snapshot to be successfully built and the test case 
to be executed, thus reenabling the ability to use automated bisection.

\section{Design and Implementation}
\label{methodology}

\begin{figure}
\centering
\begin{tikzpicture}[node distance=2cm]
    \node (start)[startstop] { Start };
    \node (copy0)[io, below of=start, yshift=-0.5cm] { Copy project into workspace };
    \node (mark)[process, below of=copy0] { Mark reachable declarations };
    \node (sweep)[process, below of=mark] { Determine reachability from global context };
    \node (remove)[process, below of=sweep] { Remove/Dummy unused declarations };
    \node (declremoved)[decision, below of=remove, aspect=3] { Declarations removed? };
    \node (copy1)[io, right of=remove, xshift=5cm, yshift=1cm] { Copy minimized project into workspace };
    \node (end)[startstop, below of=declremoved, yshift=-0.5cm] { End };

    \draw[arrow] (start) -- (copy0);
    \draw[arrow] (copy0) -- (mark);
    \draw[arrow] (mark) -- (sweep);
    \draw[arrow] (sweep) -- (remove);
    \draw[arrow] (remove) -- (declremoved);
    \draw[arrow] (declremoved) -- node[anchor=east] { No } (end);
    \draw[arrow] (declremoved) -| node[anchor=west] { Yes } (copy1);
    \draw[arrow] (copy1) |- (mark);
\end{tikzpicture}
\caption{Flow graph for the process of test case minimization.}
\label{fig:flow-graph}
\end{figure}
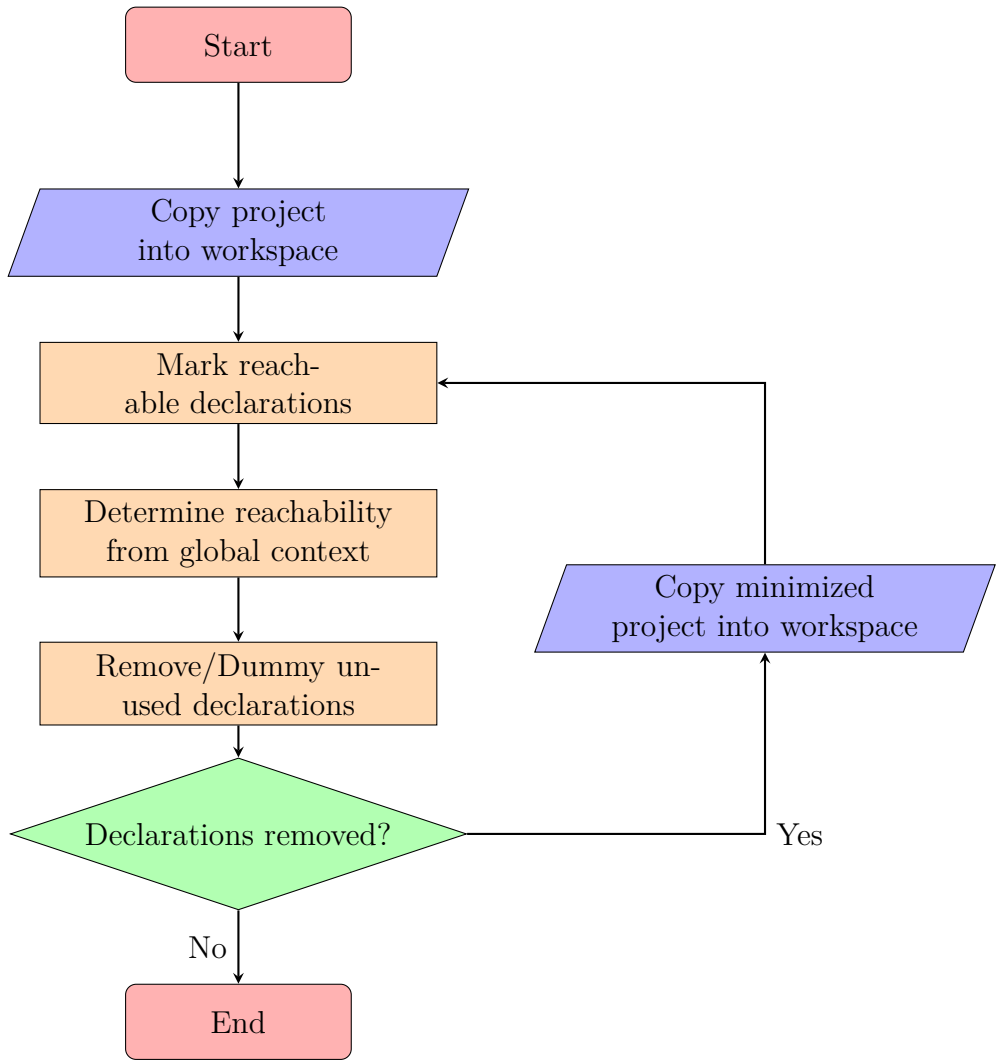

This paper proposes a novel technique that performs \textit{test dependency minimization}, which aims to minimize the 
classes and methods required to execute a test case. Since the minimization aims to repair the compilation of broken 
snapshots, minimization is performed on the source-code level as opposed to the bytecode level. Figure 
\ref{fig:flow-graph} demonstrates the high-level view of the minimization process.

\subsection{Reachability}

As the proposed technique operates on source code and utilizes static analysis, this section discusses the rules of 
reachability as used by the technique.

\textbf{For the proposed technique, we say that a program declaration is reachable if the declaration must be required 
in the successful compilation of the project, or the declaration may be required in the correct execution of the 
project.}

\subsubsection{Entrypoint}
\label{methodology:entrypoint}

Entrypoints refer to more locations where the software can begin execution. In a bisection context, entrypoints are 
generally test cases, as well as methods used to set up or clean up the test environment. To handle methods that are 
invoked internally by the test framework and are otherwise statically unreachable, we formulate a list of methods and 
constructors which, if a test class or any method within the test class is deemed to be an entrypoint, is to be marked 
as an entrypoint as well. This is also done for superclasses of the entrypoint test class, as these methods are invoked 
hierarchically.

All entrypoints must be reachable, as these methods are where the program begins executing from.

Multiple entrypoints can be specified to perform minimization on multiple test cases within a single minimization 
invocation, reducing the time needed for minimizing and running multiple test cases or test suites.

\subsubsection{Direct Reachability}
\label{directly:static_analysis}

A program declaration is directly reachable if it is a program entrypoint, or if it is required in the compilation or 
execution of another directly reachable declaration. In other words, a directly reachable declaration is 
unconditionally required in the compilation and/or execution of the snapshot.

The most granular level for which reachability can be determined using static analysis is class members. This is 
because the possible execution of statements depends on the actual value of all subexpressions within the statement, 
which may be impossible to accurately determine if the value is dependent on external input, and may cause state 
explosion for methods that are invoked in many locations.

\subsubsection{Transitive Reachability}
\label{transitive:static_analysis}

A program declaration is transitively reachable if it may be needed during compilation or execution, but its 
reachability cannot be statically determined due to the use of dynamic properties such as branching and dynamic method 
lookup. 

\begin{lstlisting}[
    language=Java,
    label={lst:transitive-override},caption={Example of transitive reachability by overriding methods.}
]
public class I {
    public int getInt() { return 0; }
}

public class A extends I {
    public int getInt() { return 1; }
}

public class B extends I {
    public int getInt() { return 2; }
}

public class Main {

    public static void main(String[] args) {
        I i = args.length > 1 ? new A() : new B();
        int result = i.getInt();
    }
}
\end{lstlisting}

Referring to Listing \ref{lst:transitive-override}, and assuming that \texttt{main} is the sole entrypoint, when 
statically analyzing the expression \texttt{i.getInt()}, it is definitively known that \texttt{I.getInt()} is directly 
reachable since it is the static target of the method call. However, the actual method invoked by this method call is 
unknown, as it could be the method in \texttt{A} or \texttt{B}.

The four cases where declarations are transitively reachable are

\begin{itemize}
\item \textbf{Transitive Constructors for Class} 

All constructors in a class are marked as transitively reachable. This is because if the superclass does not have a 
no-argument constructor, at least one constructor in the class needs to be kept for compilation to succeed.

\item \textbf{Transitive Constructors for Subclass} 

All constructors in a class that may be used to delegate a \texttt{super} explicit constructor invocation statement 
are marked as transitively reachable so that constructors in the subclass can use the constructor for delegation.

\item \textbf{Dynamic Lookup Targets} 

Methods that are potential targets of dynamic method lookup (hereinafter referred to as \textit{dynamic lookup 
targets}) are marked as transitive reachable. This is because while the full set of dynamic lookup targets can be known 
given the whole program, this set can be reduced if information such as whether a class may be instantiated can be 
used. This optimization will be further discussed in Section \ref{opt:mindispatchset}.
    
\item \textbf{Library Call Targets} 

In some cases where a class overrides methods from a supertype in a library, the overridden method can still be 
reachable even if no static call resolves to it. This is because while the library can only statically invoke methods 
that are visible to the library, the method can be still invoked by dynamic method lookup.
\end{itemize}

\subsubsection{Reachability Reasons}
\label{reachability-reasons}

As mentioned in Sections \ref{directly:static_analysis} and \ref{transitive:static_analysis}, there are two main 
categories of reachability. We can further categorize these types of reachability into specific reasons, as different 
types of reasons demand different ways to determine whether a program declaration is reachable relative to other 
declarations.

The list of reasons why a declaration may be needed for compilation and/or execution can be grouped into one of four 
categories:

\begin{itemize}
\item \textbf{Referenced by Symbol:} In general, if a symbol identifier is present in the source code, the declaration 
of the symbol must be present for the compilation to succeed.

\item \textbf{Dynamic Call Target:} Java invokes non-static methods using dynamic method lookup by default, meaning 
that 
the static target of the method or any overriding method may be executed depending on the concrete type of the operand. 
As such, these overriding methods may be needed in the execution of the test case.

\item \textbf{Constructor Delegation:} Java requires constructors to invoke any superclass constructor to enforce the 
proper creation of a derived class instance. Therefore, if the superclass does not declare any no-argument constructor, 
its derived classes must declare at least one constructor which invokes a superclass constructor for successful 
compilation.

\item \textbf{Parent Construct:} This concerns declarations that can contain nested declarations, such as parents of 
nested classes. The parent class of a nested class is needed for compilation as it provides the scope for the nested 
class.
\end{itemize}

\subsubsection{Reachability Graph}

When the reachability reasons for all relevant program declarations are determined, we can determine whether each
declaration should be retained, dummied, or removed. To do so, we use the observation that the relationship between
declarations and reachability reasons can be formulated into a directed graph, where declarations are nodes and 
reachability reasons are directed edges. We can then use depth-first search to follow each edge until we can determine 
whether any dependent node is reachable, which is finally used to determine whether the declaration will be retained, 
dummied, or removed in a manner as described in Section \ref{sweep-phase}.

One thing to note is that the graph may be cyclic as declarations may be dependent on one another, such as when two 
methods recursively invoke one other. In those cases, the edge that forms the cycle will not be traversed, effectively 
treating nodes within a cycle as reachable only if any of the nodes are reachable due to an edge that is not involved 
in the cycle.

\subsection{Minimization Phases}
\label{minimization-phases}

Transitively reachable declarations require the context of the entire program to accurately determine whether it is 
reachable. To address this issue, we propose a two-phase strategy that separates the determination of reachability into 
two parts, the \textit{mark phase} and the \textit{sweep phase}.

\subsubsection{Mark Phase}
\label{mark-phase}

The mark phase is responsible for marking the reachability type for each program declaration, based on the rules 
established in Section \ref{reachability-reasons}. The mark phase begins searching for identifiers from all entrypoints 
of the program, iteratively adding declarations until no new declarations are found. Moreover, this phase also 
conservatively includes all declarations that are transitively reachable but may not be necessary for compilation or 
execution to succeed.

When each declaration is processed, the reason(s) for why the declaration is reachable is also stored alongside the 
declaration.

\subsubsection{Sweep Phase}
\label{sweep-phase}

After all reachable program declarations are marked, the sweep phase is responsible for determining whether each 
transitively reachable declaration is necessary to maintain the compilability and execution correctness of the program. 
The sweep phase considers the following information when determining the necessity of a program declaration.

\begin{itemize}
\item \textbf{Declaration Reachability}

As described in Section \ref{direct-transitive-reachability}, a directly reachable declaration is always needed for 
compilation, because these declarations are either likely to be executed, or used in declarations or executable code. 
Moreover, if a transitively reachable declaration is referenced by a directly reachable declaration, it means that the 
presence of the transitively reachable declaration is required for the directly reachable declaration to successfully 
compile and/or execute.

\item \textbf{Container Reachability}

If the type or body declaration containing the declaration is not reachable, it means that no declaration requires 
this for compilation or execution, and is therefore removable.

\item \textbf{Class Instantiation}

Non-static member declarations require an instance of the type to operate upon. Therefore, if the declaring class is 
never instantiated, there are no instances to invoke non-static members on, and thus all non-static members are 
removable.

\item \textbf{Non-Static Method Usage}

A non-static method in a class with subclasses cannot be removed, as all subclasses that do not override the method 
will lose its implementation. A method overriding an abstract method cannot be removed either, as all concrete classes 
require an implementation for all its methods.

\item \textbf{Field Initializer}

If a field has an initializer expression, it should be retained regardless, because the initializer expression may have 
side effects that modify the program state.
\end{itemize}

For each program declaration, one of three decisions will be made depending on its necessity in compilation and 
execution.

\begin{itemize}
\item \textbf{No-Op:} No-Op retains the declaration in its entirety, except for declarations that may contain nested 
declarations. This usually applies to declarations that are directly reachable.

\item \textbf{Dummy:} Dummy only retains a subset of the declaration while still allowing the declaration to be 
referenced by name. This usually applies to declarations that are transitively reachable.

To address the issue of diagnosability as described in Section \ref{technical-challenges}, our technique supports 
injecting statements into the body of dummied methods and constructors to aid the diagnosis of unsound static analysis. 
The goal of this is to report unexpected execution as early as possible instead of relying on an upstream caller or 
assertion to catch any unexpected values. In the implementation of our technique, we opted to replace the dummy values 
with a statement that throws an \texttt{AssertionError}. When such an error is caught, the method throwing the 
assertion error can be added to the list of entrypoints to explicitly include the method and its dependencies in the 
minimized output.

\item \textbf{Remove:} Remove fully removes the declaration from the project. This usually applies to declarations that 
are not reachable.
\end{itemize}

After the sweep phase is executed, each program declaration will be transformed based on the decision made during the 
sweep phase.

\subsubsection{Example}

Using the example from Listing {\ref{lst:transitive-override}}, when using two-phase minimization, the observations 
stated in Section \ref{direct-transitive-reachability} are established in the mark phase. When the sweep phase is 
executed, the following additional observations can be made within the context of the whole program.

\begin{itemize}
\item No constructor to \texttt{I} is invoked anywhere in the program, meaning that no object with the concrete type of 
\texttt{I} will exist in the lifetime of the program.

\item Both \texttt{A} and \texttt{B} override the implementation of \texttt{I.getInt}.

\item As no instance of \texttt{I} is created, and all non-abstract subclasses of \texttt{I} provide their own 
implementation of \texttt{I.getInt}, the implementation of \texttt{I.getInt} is never used.
\end{itemize}

With the aforementioned observations established, the sweep phase can conclude that the body of \texttt{I.getInt} is no 
longer needed, and thus the method will be marked for dummying.

\subsection{Optimizations}

\subsubsection{Minimizing the Set of Dynamic Lookup Targets}
\label{opt:mindispatchset}

As mentioned in \ref{transitive:static_analysis}, including all possible targets in a dynamic lookup context can 
introduce many redundant program declarations, especially when the scope type is of a common library type such as 
\texttt{Object}. To address this problem, we implement additional inference logic to reduce redundant methods for a 
method call involving dynamic lookup.

\paragraph{Variable Usage} For expressions that reference a variable in its scope, since the entire program is known to 
the technique, the static type of all values assigned to the variable is known. Therefore, we can use this information 
to narrow the types of objects which may be assigned to the variable. This information is especially useful for 
variables declared with a type where the number of subclasses and therefore the number of methods overriding the 
top-level class is large.

\paragraph{Generics in Class Fields} While the set of all types assigned to a variable is useful in narrowing the set 
of dynamic lookup targets, this is insufficient when the variable is a field within a generic class. This is because 
these fields may be assigned any type of value as long as it satisfies the bounds of the class type parameter, but when 
a field variable is used the type parameter is replaced with a type variable specific to the context, and thus only a 
subset of assigned types to the field are valid values in the context.

\begin{lstlisting}[
    language=Java,
    label={lst:generic-class-field},
    caption={An example of minimizing types for a generic class field.}
]
public class Pair<A, B> {

    public A first;
    public B second;
}

public class Main {

    public static void Main(String[] args) {
        Pair<Set<?>, Object> pair1 = new Pair<>();
        Pair<List<?>, Object> pair2 = new Pair<>();

        pair1.first = new HashSet<Object>();
        pair2.first = new ArrayList<Object>();

        getString(pair1);
    }
}
\end{lstlisting}

Listing \ref{lst:generic-class-field} demonstrates a program containing a generic type and two instances with different 
type parameters that are instantiated in the \texttt{main} method. Since the type of \texttt{pair1.first} is both 
constrained by the type of values assigned to the field as well as the type of the variable, only 
\texttt{HashSet<Object>} would be a valid assigned type in this context.

\subsubsection{Generics Type Information Propagation}

When determining the set of possible dynamic lookup targets, determining the type of expressions in generic contexts is 
important for improving accuracy. This is because while generic parameters can be easily solved as they appear in 
the class or method header, generic type variables can be present in a subexpression, and the type of the full 
expression may be dependent on the type of the subexpression, such as in chained method calls or nested method calls.

\begin{lstlisting}[language=Java,label={lst:generic-method-chain},caption={Simplified example of chained generics in 
JacksonDatabind-1f.}]
public static Class<? extends Enum<?>> findEnumType(EnumMap<?,?> m)
{
    // ...
    return findEnumType(m.keySet().iterator().next());
}
\end{lstlisting}

Listing \ref{lst:generic-method-chain} shows a method with the expression \texttt{m.keySet().iterator().next()}, where 
each subexpression returns a generic type. We want to solve the type of this expression to find which of the four 
method 
overloads below will be selected for invocation:

\begin{itemize}
\item \texttt{findEnumType(EnumSet<?>)}
\item \texttt{findEnumType(EnumMap<?>)}
\item \texttt{findEnumType(Enum<?>)}
\item \texttt{findEnumType(Class<?>)}
\end{itemize}

A naive approach to solving the type of this expression would be to recursively inspect the type of each subexpression 
and formulate the type of the full expression. However, using such a naive approach causes some generic constraints to 
be lost during the solving process. Therefore, we implement a custom generics solver that takes into account all 
generic constraints declared by the class, method, and variable to more accurately solve the type of the expression.

\begin{table}
\centering
\begin{tabular}{ ||c || c | c ||}
    \hline\hline
    Subexpression & Naive Type & Solved Type \\
    \hline\hline
    \texttt{m} & \texttt{EnumMap<?, ?>} & \texttt{EnumMap<?, ?>} \\
    \texttt{keySet()} & \texttt{Set<?>} & \texttt{Set<? extends Enum<?>>} \\
    \texttt{iterator()} & \texttt{Iterator<?>} & \texttt{Iterator<? extends Enum<?>>} \\
    \texttt{next()} & \texttt{? extends Object} & \texttt{Enum<?>} \\
    \hline\hline
\end{tabular}
\caption{Comparison between solving generic typed expressions with and without Generic Type Information Propagation.}
\label{table:gtip}
\end{table}

As seen from Table \ref{table:gtip}, the approach using the custom solver allows more generic information to be 
retained and can infer a more specific type than the naive approach. As such, when finding a candidate method 
for \texttt{findEnumType}, the custom solver achieves greater accuracy than the naive approach.

\subsection{Multiple Passes}

In the optimization technique which minimizes the set of dynamic lookup targets described in Section 
\ref{opt:mindispatchset}, we find all initialization and assignment expressions to a variable to determine the set of 
types that the variable may store. When constructors and methods are being removed in the Sweep Phase, this may remove 
expressions which assign to the variable, meaning that the variable may be assigned to fewer values and types. This, 
in turn, causes fewer methods to be identified as viable dynamic method lookup candidates, which allows more methods to 
be dummied or removed.

Therefore, we extend the algorithm to perform multiple passes of the mark-sweep process to utilize an updated context 
after some methods have been removed, which opens up new opportunities to identify unreachable methods. Since the 
minimization process never introduces new symbols, the mark-sweep process can be repeated until no more symbols are 
removed, in which case we consider the minimization to have reached convergence.

\subsubsection{Example}

Listing \ref{lst:two-phase-mod} shows an example modified from Listing \ref{lst:transitive-override}. The 
execution behavior of the program remains unchanged; However, the behavior of the technique is changed due to the 
following reasons:

\begin{lstlisting}[
    language=Java,
    caption={Example for illustrating multiple passes.}
    ,label={lst:two-phase-mod}
]
public class A {}
public class B {}

public abstract class Abstract {
    abstract void f();
}
public class AbstractImpl1 extends Abstract {
    abstract void f() {
        new A();
    }
}
public class AbstractImpl2 extends Abstract {
    abstract void f() {
        new B();
    }
}

public class Main {
    public static void foo(Abstract obj) {
        obj.f();
    }

    public static void main(String[] args) {
        foo(new AbstractImpl1());
    }
}
\end{lstlisting}

\begin{itemize}
\item Since the set of possible types of the parameter is assumed to be \texttt{? extends Abstract}, the mark phase 
must consider both \texttt{AbstractImpl1} and \texttt{AbstractImpl2} as reachable, which follows that \texttt{A} and 
\texttt{B} are also reachable.

\item During the sweep phase, it is known that only \texttt{AbstractImpl1} is instantiated in the context of the whole 
program, meaning that the body of \texttt{AbstractImpl2.f} can never be executed. Therefore, \texttt{AbstractImpl2.f} 
will be marked as dummiable, but \texttt{B} is still marked for retention as \texttt{AbstractImpl2.f} is not marked for 
removal.
\end{itemize}

The result is that while \texttt{AbstractImpl2.f} is dummied, \texttt{B} is not removed, leaving room for further 
minimization. If a second pass is executed, \texttt{B} will be successfully marked for removal, which removes an extra 
class from the minimized program.

\section{Evaluation}
\label{eval}

This section presents the evaluation of test dependency minimization. Specifically, we would like to answer the 
following research questions:

\begin{itemize}
\item \textbf{RQ1 (Causes of Build Failures):} What are the reasons snapshots cannot be compiled using newer versions 
of 
Java?

\item \textbf{RQ2 (Effectiveness on Build Repair):} To what extent is our technique able to automatically repair 
compilation errors?

\item \textbf{RQ3 (Accuracy of Minimization):} To what extent does our technique accurately minimize the number of 
classes and methods used in the snapshot?

\item \textbf{RQ4 (Technique Overhead):} How much overhead does our technique introduce?
\end{itemize}

RQ1 aims to provide an updated context for our tool by re-evaluating the causes of build failures using newer versions 
of JDK. RQ2 and RQ3 aim to evaluate the effectiveness of our technique by evaluating its ability to repair and the 
accuracy of retained program declarations compared to a class-granular minimization approach. RQ4 aims to evaluate the 
overhead of our tool by comparing the runtime with the full compilation of the snapshot.

\subsection{Evaluation Subjects}

\paragraph{Subject Selection} We use Defects4J to evaluate all research questions.

\paragraph{Evaluation Environment} The machine specifications used to evaluate the tool are as follows (unless 
otherwise specified):

\begin{itemize}
\item 128-thread AMD Ryzen Threadripper PRO 3995WX (48 threads allocated to the JVM process)

\item 512GB RAM (32GB allocated to the JVM process)

\item Docker 24.0.1

\item OpenJDK 1.8.0\_352, 17.0.5

\item CentOS Stream 8
\end{itemize}

\subsection{RQ1: Causes of Build Failures}

\paragraph{Experiment Setup} To understand the reasons why the compilation of snapshots would fail using current 
toolchains, we took all bugs in Defects4J and recompiled them using Java 8 and 17 respectively. These Java versions are 
selected because they are the oldest and newest currently-supported LTS versions of Java respectively, meaning that 
developers are more likely to target these versions of Java \cite{oraclejavaltsblog}. Moreover, we choose Java 17 
instead of the more popular Java 11 because we want to investigate the worst-case scenario when a bisection needs to be 
performed on a project that targets the latest version of Java.

All Defects4J bugs are recompiled under each of the following four configurations.

\begin{enumerate}
\item \textbf{JDK 8, Source Level 1.7:} This configuration is the default environment provided by Defects4J, except 
that 
Java 8 standard library classes will be used instead.

\item \textbf{JDK 8, Source Level 1.8:} This configuration simulates the compilation of snapshots using JDK 8 and Java 
8 
language features.

\item \textbf{JDK 17, Source Level 1.7:} This configuration simulates the compilation of snapshots using JDK 17 with 
maximum compatibility with language features in Java 7. Note that this configuration is deprecated in Java 17; The 
oldest supported source level is Java 8.

\item \textbf{JDK 17, Source Level 17:} This configuration simulates the compilation of snapshots using JDK 17 and Java 
17 language features.
\end{enumerate}

\paragraph{Results} The breakdown of compilation error reasons for each Java version is shown in Tables 
\ref{table:rq1jdk8java7}-\ref{table:rq1jdk17java17}. The reasons why a bug is uncompilable under newer versions of Java 
are split into 4 categories, based on how the compilation failure is manifested and the possible solutions to fixing 
the 
issue.

\begin{itemize}
\item Changes to the Java Language (\textit{Lang-Change})

This category of bugs fails to compile under newer versions of Java because these revisions contain program constructs 
that are forbidden in the newer Java versions and will cause compilation errors, such as declaration of identifiers 
that later became keywords.

\item Changes to the Java Standard Library (\textit{Lib-Change)}

This category of bugs fails to compile under newer versions of Java because these revisions use classes or methods in 
the Java Standard Library which is changed in newer Java versions, such as the addition or removal of methods and/or 
overloads.

\item Unsupported Character Encoding (\textit{Encoding})

This category of bugs fails to compile under newer versions of Java because these revisions contain character sequences 
that are invalid under UTF-8.

\item Unsupported Tools (\textit{Tool})

This category of bugs fails to compile under newer versions of Java because these revisions use tools that do not 
support newer versions of Java.
\end{itemize}

\begin{table}
\centering
\resizebox{\textwidth}{!}{
    \begin{tabular}{ || c c | c c c c || }
        \hline\hline
        Project         & Subjects & Lang-Change & Lib-Change & Encoding  & Tool \\
        \hline\hline
        Chart           & 52       & -           & -          & -         & -    \\
        Cli             & 78       & -           & -          & -         & -    \\
        Closure         & 348      & -           & -          & -         & -    \\
        Codec           & 36       & -           & -          & 14 (39\%) & -    \\
        Collections     & 8        & -           & 4 (50\%)   & -         & -    \\
        Compress        & 94       & -           & -          & -         & -    \\
        Csv             & 32       & -           & -          & -         & -    \\
        Gson            & 36       & -           & -          & -         & -    \\
        JacksonCore     & 52       & -           & -          & -         & -    \\
        JacksonDatabind & 224      & -           & -          & -         & -    \\
        JacksonXml      & 12       & -           & -          & -         & -    \\
        Jsoup           & 186      & -           & -          & -         & -    \\
        JxPath          & 44       & -           & -          & -         & -    \\
        Lang            & 128      & 92 (72\%)   & -          & 56 (44\%) & -    \\
        Math            & 212      & -           & -          & 38 (18\%) & -    \\
        Mockito         & 76       & -           & -          & -         & -    \\
        Time            & 52       & -           & -          & -         & -    \\
        \hline
        Total           & 1670     & 92 (6\%)    & 4 (0\%)    & 108 (6\%) & -    \\
        \hline\hline
    \end{tabular}
}
\caption{Reasons for build failure by project when compiling using Java 8, source level 1.7.}
\label{table:rq1jdk8java7}
\end{table}

\begin{table}
\centering
\resizebox{\textwidth}{!}{
    \begin{tabular}{ || c c | c c c c || }
        \hline\hline
        Project         & Subjects & Lang-Change & Lib-Change & Encoding  & Tool       \\
        \hline\hline
        Chart           & 52       & -           & -          & -         & -          \\
        Cli             & 78       & -           & -          & -         & -          \\
        Closure         & 348      & -           & -          & -         & 210 (60\%) \\
        Codec           & 36       & -           & -          & 14 (39\%) & -          \\
        Collections     & 8        & 5 (63\%)    & -          & -         & -          \\
        Compress        & 94       & -           & -          & -         & -          \\
        Csv             & 32       & -           & -          & -         & -          \\
        Gson            & 36       & -           & -          & -         & -          \\
        JacksonCore     & 52       & -           & -          & -         & -          \\
        JacksonDatabind & 224      & -           & -          & -         & -          \\
        JacksonXml      & 12       & -           & -          & -         & -          \\
        Jsoup           & 186      & -           & -          & -         & -          \\
        JxPath          & 44       & -           & -          & -         & -          \\
        Lang            & 128      & 48 (38\%)   & -          & 56 (44\%) & -          \\
        Math            & 212      & -           & -          & 38 (18\%) & -          \\
        Mockito         & 76       & 46 (61\%)   & -          & -         & -          \\
        Time            & 52       & -           & -          & -         & -          \\
        \hline
        Total           & 1670     & 99 (6\%)    & -          & 108 (6\%) & 210 (13\%) \\
        \hline\hline
    \end{tabular}
}
\caption{Reasons for build failure by project when compiling using Java 8, source level 1.8.}
\label{table:rq1jdk8java8}
\end{table}

\begin{table}
\centering
\resizebox{\textwidth}{!}{
    \begin{tabular}{ || c c | c c c c || }
        \hline\hline
        Project         & Subjects & Lang-Change & Lib-Change  & Encoding  & Tool      \\
        \hline\hline
        Chart           & 52       & -           & -           & -         & -         \\
        Cli             & 80       & -           & -           & -         & -         \\
        Closure         & 348      & 94 (27\%)   & -           & -         & -         \\
        Codec           & 36       & -           & -           & 14 (39\%) & -         \\
        Collections     & 8        & -           & 8 (100\%)   & -         & -         \\
        Compress        & 94       & -           & 78 (83\%)   & -         & -         \\
        Csv             & 32       & -           & -           & -         & -         \\
        Gson            & 36       & -           & -           & -         & -         \\
        JacksonCore     & 52       & -           & -           & -         & -         \\
        JacksonDatabind & 224      & -           & 198 (88\%)  & -         & -         \\
        JacksonXml      & 12       & -           & 8 (67\%)    & -         & -         \\
        Jsoup           & 186      & -           & -           & -         & -         \\
        JxPath          & 44       & -           & -           & -         & -         \\
        Lang            & 128      & 92 (72\%)   & -           & 56 (44\%) & -         \\
        Math            & 212      & -           & 212 (100\%) & -         & -         \\
        Mockito         & 76       & -           & -           & -         & 30 (39\%) \\
        Time            & 52       & -           & -           & -         & -         \\
        \hline
        Total           & 1670     & 186 (11\%)  & 504 (30\%)  & 70 (4\%)  & 30 (2\%)  \\
        \hline\hline
    \end{tabular}
}
\caption{Reasons for build failure by project when compiling using Java 17, source level 1.7.}
\label{table:rq1jdk17java7}
\end{table}

\begin{table}
\centering
\resizebox{\textwidth}{!}{
    \begin{tabular}{ || c c | c c c c || }
        \hline\hline
        Project         & Subjects & Lang-Change & Lib-Change  & Encoding  & Tool      \\
        \hline\hline
        Chart           & 52       & -           & -           & -         & -         \\
        Cli             & 80       & -           & -           & -         & -         \\
        Closure         & 348      & 94 (27\%)   & -           & -         & -         \\
        Codec           & 36       & -           & -           & 14 (39\%) & -         \\
        Collections     & 8        & -           & 8 (100\%)   & -         & -         \\
        Compress        & 94       & -           & 78 (83\%)   & -         & -         \\
        Csv             & 32       & -           & -           & -         & -         \\
        Gson            & 36       & -           & -           & -         & -         \\
        JacksonCore     & 52       & -           & -           & -         & -         \\
        JacksonDatabind & 224      & -           & 198 (88\%)  & -         & -         \\
        JacksonXml      & 12       & -           & 8 (67\%)    & -         & -         \\
        Jsoup           & 186      & -           & -           & -         & -         \\
        JxPath          & 44       & -           & -           & -         & -         \\
        Lang            & 128      & 92 (72\%)   & -           & 56 (44\%) & -         \\
        Math            & 212      & -           & 212 (100\%) & -         & -         \\
        Mockito         & 76       & -           & -           & -         & 30 (39\%) \\
        Time            & 52       & -           & -           & -         & -         \\
        \hline
        Total           & 1670     & 186 (11\%)  & 504 (30\%)  & 70 (4\%)  & 30 (2\%)  \\
        \hline\hline
    \end{tabular}
}
\caption{Reasons for build failure by project when compiling using Java 17, source level 17.}
\label{table:rq1jdk17java17}
\end{table}

Firstly, we can observe that Java 8 contains the least number of compilation errors, which matches the data collected 
by previous empirical studies. We also observe that Java 17 contains the greatest number of compilation errors, which 
can be explained by the number of changes between these two Java versions.

Secondly, when bumping the source level from 1.7 to the level supported by the compiler, we can observe that the number 
of build failures increases for JDK 8, whereas there is no change for JDK 17. This can be explained by Java 8 
containing changes to how generic types are computed and how methods are resolved, causing new compilation errors to be 
emitted. As for Java 17, this lack of change can be attributed to that the language features and library APIs used by 
the subjects are so old that support for them has been removed regardless of the chosen language level.

Finally, we can observe that library changes and language changes account for the greatest increase in compilation 
errors between Java 8 and 17. This can be explained by the number of language changes and API changes between Java 8 
and 17. One interesting note is that some revisions manifest different categories of compilation errors when compiling 
under different JDKs.

\subsection{RQ2: Effectiveness on Build Repair}
\label{rq2}

We propose two alternative techniques to act as a ground truth and baseline respectively.

\paragraph{Ground Truth} The ground truth is provided by coverage-based minimization. Coverage-based minimization 
utilizes coverage data to determine the reachability of each program declaration and remove all unreachable 
declarations. Since coverage data is collected by executing the program, the coverage data collected for all methods 
and classes is representative of the program components necessary at runtime for a given entrypoint.

The coverage data is collected by compiling the snapshot using a supported JDK version. This coverage data is then used 
to determine whether each declaration and statement in the snapshot is reachable. If not, the unreachable program 
component will either be dummied or removed, depending on whether it is needed in the compilation of other reachable 
program components.

The ground truth will be used to determine whether each subject can be successfully compiled after minimization, since 
if a compilation error still occurs after coverage-based minimization, a coarser-grained algorithm is unlikely to be 
able to perform the repair either.

We use two existing coverage tools to collect coverage data for each subject: Cobertura 
\footnote{https://github.com/cobertura/cobertura} and Jacoco \footnote{https://www.eclemma.org/jacoco/}. The reason why 
two coverage tools are selected is that previous trials have shown that Cobertura and Jacoco inject instrumentation 
statements in different locations of the bytecode, which causes inconsistencies when determining whether a statement is 
covered.

\paragraph{Baseline} The baseline for all subjects is provided by class-granular minimization. Like the proposed 
technique, class-granular minimization uses static analysis to find all reachable program declarations from the program 
entrypoint, but with the difference that only the reachability of classes are determined, meaning that all members are 
included for a reachable class, regardless of the reachability of each member.

We chose this technique as our baseline because this is the most commonly used technique for statically analyzing 
declaration dependencies at both source-level and bytecode-level.

\paragraph{Experiment Setup} To evaluate the effectiveness of our technique, we use a subset of test cases in the 
Defects4J dataset which satisfies the following criteria:

\begin{itemize}
    \item The bug cannot be directly compiled under a newer version of Java.
    \item The test case exposes the bug.
    \item The test case can be compiled and executed after running minimization using coverage data.
\end{itemize}

All test cases which satisfy the above criteria are collected as subjects for evaluation. Note that one Defects4J bug 
may provide more than one subject to the evaluation, as there may be multiple triggering test cases for a single 
Defects4J bug.

For each collected test case, we run our technique and the baseline under two JDK environments.

\begin{itemize}
\item \textbf{JDK 8, Source Level 1.7:} This environment is equivalent to Configuration 1 in RQ1 and is used to 
simulate 
build repair in an environment using the oldest supported compiler and an older source version to maximize 
compatibility with older snapshots.

\item \textbf{JDK 17, Source Level 17:} This environment is equivalent to Configuration 4 in RQ1 and is used to 
simulate 
an environment where a bisection process covers snapshots that use any version of Java between Java 1.1 and 17.
\end{itemize}

After performing minimization, RQ2 will be evaluated using the following two metrics.

\begin{itemize}
\item \textbf{Compilability:} This checks whether the resultant minimized project is compilable using a Java compiler 
version and source level.

\item \textbf{Test Match} This checks whether the execution result of the test case matches the expected result. The 
expected result is determined as follows:

\begin{itemize}
\item If the checked-out project version is the buggy version, and the test case is a triggering test, the test case is 
expected to fail.

\item Otherwise, the test case is expected to pass.
\end{itemize}
\end{itemize}

\begin{table}
\centering
\resizebox{\textwidth}{!}{
    \begin{tabular}{ || c  c | c  c | c  c || }
        \hline\hline
        \multirow{2}{*}{Project} & \multirow{2}{*}{Total Count} & \multicolumn{2}{|c|}{Baseline} & 
\multicolumn{2}{|c||}{Member-Granular} \\
        & & Compilable & Test Match & Compilable & Test Match \\
        \hline\hline
        Codec       & 20  & 20 (100\%) & 20 (100\%) & 20 (100\%)  & 20 (100\%) \\
        Collections & 4   & 4 (100\%)  & 4 (100\%)  & 4 (100\%)   & 2 (50\%)   \\
        Lang        & 67  & 64 (96\%)  & 64 (96\%)  & 67 (100\%)  & 67 (100\%) \\
        Math        & 39  & 39 (100\%) & 39 (100\%) & 39 (100\%)  & 29 (74\%)  \\
        \hline\hline
        Total       & 130 & 127 (98\%) & 127 (98\%) & 130 (100\%) & 118 (91\%) \\
        \hline\hline
    \end{tabular}
}
\caption{Results of minimization on subjects compiled using Java 8, source level 1.7.}
\label{table:rq2-jdk8-source1.7}
\end{table}

\begin{table}
\centering
\resizebox{\textwidth}{!}{
    \begin{tabular}{ || c  c | c  c | c  c || }
        \hline\hline
        \multirow{2}{*}{Project} & \multirow{2}{*}{Total Count} & \multicolumn{2}{|c|}{Baseline} & 
\multicolumn{2}{|c||}{Member-Granular} \\
        & & Compilable & Test Match & Compilable & Test Match \\
        \hline\hline
        Closure         & 511 & 511 (100\%) & 511 (100\%) & 511 (100\%) & 468 (92\%) \\
        Codec           & 20  & 20 (100\%)  & 20 (100\%)  & 20 (100\%)  & 20 (100\%) \\
        Collections     & 7   & 3 (43\%)    & 3 (43\%)    & 7 (100\%)   & 4 (57\%)   \\
        Compress        & 103 & 103 (100\%) & 103 (100\%) & 103 (100\%) & 94 (91\%)  \\
        JacksonDatabind & 219 & 199 (91\%)  & 197 (90\%)  & 219 (100\%) & 155 (71\%) \\
        JacksonXml      & 12  & 0 (0\%)     & 0 (0\%)     & 12 (100\%)  & 8 (67\%)   \\
        Lang            & 12  & 12 (100\%)  & 12 (100\%)  & 12 (100\%)  & 12 (100\%) \\
        Mockito         & 67  & 67 (100\%)  & 36 (54\%)   & 67 (100\%)  & 35 (52\%)  \\
        \hline\hline
        Total           & 951 & 915 (96\%)  & 882 (93\%)  & 951 (100\%) & 796 (84\%) \\
        \hline\hline
    \end{tabular}
}
\caption{Results of minimization on subjects compiled using Java 17, source level 17.}
\label{table:rq2-jdk17-source17}
\end{table}

\paragraph{Results} The result of RQ2 compiled using Java 8 with source level 1.7 and Java 17 with source level 17 are 
shown in Tables {\ref{table:rq2-jdk8-source1.7}} and {\ref{table:rq2-jdk17-source17}} respectively. Our technique can 
restore compilability to all subjects, and a majority of subjects also match the expected test result, indicating that 
the technique is very effective.

When compared to the baseline, the data shows that the baseline is unable to restore compilability to some subjects 
from Collections, JacksonDatabind, and JacksonXml. We performed a manual review on these subjects and can conclude that 
due to the coarse granularity of the baseline algorithm, some unreachable methods containing compilation errors will be 
included because its container class is reachable, therefore failing to fix the compilation error.

However, the data also shows that the baseline can replicate test results more often than our technique. We also 
manually review these subjects and can conclude that all of these subjects invoke Java Reflection APIs during runtime, 
which introduces unsoundness into our technique. This is further discussed in Section \ref{discussion}.

\subsubsection{Usage in Bisection}
\label{eval:bisect}

\mod{Let us illustrate the use of our tool for automated bisection using a bug from the Defects4J 
corpus. It is a bug designated \texttt{LANG-747} in the Apache Commons Lang project.}

The statement causing the compilation error and issue with compiling the revision in Java 8 is 
shown in Listing \ref{lst:lang1-err-stmt} and \ref{lst:lang1-err-msg} respectively.

\begin{lstlisting}[numbers=none,language=Java,label={lst:lang1-err-stmt},caption={The statement causing the 
compilation error.}]
final Integer max = TypeUtilsTest.stub();
\end{lstlisting}

\begin{lstlisting}[numbers=none,label={lst:lang1-err-msg},caption={The compilation error message while compiling 
\texttt{LANG-747} using Java 8.}]
src/test/java/org/apache/commons/lang3/reflect/TypeUtilsTest.java:[524,47] incompatible types: inferred type does not conform to upper bound(s)
     inferred: G
     upper bound(s): java.lang.Comparable<G>,java.lang.Integer
\end{lstlisting}

\mod{The error message indicates that the compilation failed due to incompatible generic types. This is because Java 8 
improved the type inference algorithm to support target-typing, however, it also means that the declared variable type 
will not be used in determining the type of the expression.}

We select this bug because it demonstrates an instance where the first bad revision of the bug targets a different 
version of Java than the latest version of Java when the bug is first discovered. While developers can use any 
supported version of JDK to compile the project revision as JDK 8 supports compiling Java 6 sources, \mod{the Java 
compiler does not fully support falling back to the legacy type inference behavior.}

We incorporate our tool into the bisection process by writing a script that performs the following logic and running 
the bisection using \texttt{git bisect run}:

\begin{enumerate}
\item Extract a patch containing only the regression test case for the bug and apply it to the currently tested 
revision.

\item Try to compile the sources. If the compilation succeeds, run the patched test case and exit the script.

\item Extract the classpath of the project and use that to run our minimization tool. Delete all source files in the 
project and replace them with our minimized sources.

\item Try to compile the minimized sources. If the compilation succeeds, run the patched test case and exit the script. 
Otherwise, mark the revision as skipped.
\end{enumerate}

The machine specifications used for bisection are as follows:

\begin{itemize}
\item 24-thread AMD Ryzen 9 5900X (24 threads allocated to the JVM process)

\item 64GB RAM (16GB allocated to the JVM process)

\item OpenJDK 1.8.0\_382

\item Arch Linux
\end{itemize}

\mod{We decided to limit the bisection to all commits utilizing JUnit 4, as the test case is written using JUnit 4 
APIs, and our goal is to demonstrate the usefulness of our tool with regards to automatic bisection. The above 
bisection process spans 268 revisions, and using automated bisection with the aid of our tool, we located 
\texttt{fe235bb} as the first buggy revision. This commit is the revision immediately after the refactoring to JUnit 4, 
likely meaning that the bug exists before the refactoring. The bisection script is executed 10 times, 
and the entire bisection process takes 100 seconds, averaging 10 seconds per revision.}

\mod{We repeated this process with other bugs in the Apache Commons Lang project of the Defects4J bug corpus, and the 
successfully bisected bugs are listed in Table \ref{table:rq2-bisect-live}. Note that Lang-2 is a deprecated bug, which 
was not collected by the Defects4J corpus.}

\begin{table}
\centering
\resizebox{\textwidth}{!}{
\begin{tabular}{ || c  c | c  c  c || }
    \hline\hline
    Defects4J ID & Bug ID & No. Revisions & Bisection Time (s) & Tested Revisions \\
    \hline\hline
    Lang-1 & \texttt{LANG-747} & 268 & 100 & 10 \\
    Lang-3 & \texttt{LANG-693} & 260 & 113 & 12 \\
    Lang-4 & \texttt{LANG-882} & 98  & 85  & 8  \\
    Lang-5 & \texttt{LANG-865} & 66  & 71  & 7  \\
    Lang-6 & \texttt{LANG-857} & 60  & 87  & 7  \\
    Lang-7 & \texttt{LANG-822} & 57  & 73  & 7  \\
    Lang-8 & \texttt{LANG-818} & 26  & 70  & 6  \\
    \hline\hline
\end{tabular}
}
\caption{List of successfully bisected bugs from Apache Commons Lang of Defects4J.}
\label{table:rq2-bisect-live}
\end{table}

\mod{The experimental results demonstrate that our tool can facilitate automated bisection for uncompilable projects.}

\mod{We have also attempted to use bugs from other projects as subjects to this demonstration, but the bisection failed 
because either the test case patch cannot be applied to older revisions, or not enough revisions can be successfully 
repaired by our technique to enable the use of automated bisection. This will be further discussed in Section 
\ref{discussion}.}

\mod{We would also like to note that while Defects4J stores the good and bad revisions of each bug, the bad revision is 
either the revision before the good revision or the revision of the last stable release containing the bug. As such, 
we are unable to use Defects4J as a ground truth to compare whether our bisection result is accurate.}

\subsection{RQ3: Accuracy of Minimization}
\label{rq3-result}

\paragraph{Experiment Setup} To investigate the accuracy of minimization, we select all subjects in RQ2 that are (1) 
taken from the fixed version of bugs, and (2) able to successfully and correctly execute using both the ground truth 
and our technique, as buggy versions of subjects can fail a test case in different ways. As explained in Section 
\ref{rq2}, coverage-based technique accurately represents the program components required for the reproduction of a test 
case, hence we will continue to use it as the ground truth for this RQ. For each subject, we output the list of all 
retained classes and methods from both the coverage-based and static analysis-based techniques, and we evaluate the 
accuracy, precision, recall, and F-1 score metrics of our technique against the ground truth. In the context of this 
research question, false-positive declarations are those that should be unreachable but are retained, whereas 
false-negative declarations are those that should be reachable but are not retained.

\paragraph{Results} The statistics obtained by subjects using Java 8 and source level 1.7 are shown in Tables 
\ref{table:rq3-jdk8-java7-class-success} and \ref{table:rq3-jdk8-java7-method-success}, whereas results obtained by 
subjects using Java 17 and source level 17 are shown in Tables \ref{table:rq3-jdk17-java17-class-success} and 
\ref{table:rq3-jdk17-java17-method-success}.

\begin{table}
\centering
\resizebox{\textwidth}{!}{
    \begin{tabular}{ || c  c | c c c c c c || }
        \hline\hline
        Project & Total Count & FPR & FNR & Accuracy & Precision & Recall & F1 \\
        \hline\hline
        Codec & 10 & 0.012 & 0.114 & 0.973 & 0.936 & 0.886 & 0.902 \\
        Lang  & 29 & 0.002 & 0.179 & 0.992 & 0.957 & 0.821 & 0.864 \\
        Math  & 13 & 0.004 & 0.116 & 0.991 & 0.844 & 0.884 & 0.856 \\
        \hline\hline
        Total & 52 & 0.005 & 0.151 & 0.988 & 0.925 & 0.849 & 0.869 \\
        \hline\hline
    \end{tabular}
}
\caption{Average class accuracy of minimization on correctly executed subjects compiled using JDK 8 at source level 
1.7.}
\label{table:rq3-jdk8-java7-class-success}
\end{table}

\begin{table}
\centering
\resizebox{\textwidth}{!}{
    \begin{tabular}{ || c  c | c c c c c c || }
        \hline\hline
        Project & Total Count & FPR & FNR & Accuracy & Precision & Recall & F1 \\
        \hline\hline
        Codec & 10 & 0.004 & 0.002 & 0.996 & 0.896 & 0.998 & 0.943 \\
        Lang  & 29 & 0.002 & 0.043 & 0.998 & 0.860 & 0.957 & 0.887 \\
        Math  & 13 & 0.003 & 0.054 & 0.996 & 0.562 & 0.946 & 0.687 \\
        \hline\hline
        Total & 52 & 0.003 & 0.038 & 0.997 & 0.792 & 0.962 & 0.847 \\
        \hline\hline
    \end{tabular}
}
\caption{Average method accuracy of minimization on correctly executed subjects compiled using JDK 8 at source level 
1.7.}
\label{table:rq3-jdk8-java7-method-success}
\end{table}

\begin{table}
\centering
\resizebox{\textwidth}{!}{
    \begin{tabular}{ || c  c | c c c c c c || }
        \hline\hline
        Project & Total Count & FPR & FNR & Accuracy & Precision & Recall & F1 \\
        \hline\hline
        Closure         & 90  & 0.364 & 0.145 & 0.715 & 0.580 & 0.855 & 0.639 \\
        Codec           & 10  & 0.012 & 0.114 & 0.973 & 0.936 & 0.886 & 0.902 \\
        Compress        & 43  & 0.054 & 0.249 & 0.928 & 0.796 & 0.751 & 0.727 \\
        JacksonDatabind & 16  & 0.111 & 0.061 & 0.892 & 0.417 & 0.940 & 0.570 \\
        JacksonXml      & 2   & 0.029 & 0.000 & 0.974 & 0.719 & 1.000 & 0.835 \\
        Lang            & 5   & 0.001 & 0.061 & 0.996 & 0.988 & 0.939 & 0.960 \\
        Mockito         & 2   & 0.000 & 0.000 & 1.000 & 1.000 & 1.000 & 1.000 \\
        \hline\hline
        Total           & 168 & 0.221 & 0.156 & 0.816 & 0.660 & 0.844 & 0.687 \\
        \hline\hline
    \end{tabular}
}
\caption{Average class accuracy of minimization on correctly executed subjects compiled using JDK 17 at source level 
17.}
\label{table:rq3-jdk17-java17-class-success}
\end{table}

\begin{table}
\centering
\resizebox{\textwidth}{!}{
    \begin{tabular}{ || c  c | c c c c c c || }
        \hline\hline
        Project & Total Count & FPR & FNR & Accuracy & Precision & Recall & F1 \\
        \hline\hline
        Closure         & 90  & 0.218 & 0.041 & 0.796 & 0.291 & 0.959 & 0.426 \\
        Codec           & 10  & 0.004 & 0.002 & 0.996 & 0.896 & 0.998 & 0.943 \\
        Compress        & 43  & 0.056 & 0.101 & 0.945 & 0.543 & 0.899 & 0.616 \\
        JacksonDatabind & 16  & 0.362 & 0.015 & 0.655 & 0.122 & 0.985 & 0.216 \\
        JacksonXml      & 2   & 0.218 & 0.000 & 0.818 & 0.451 & 1.000 & 0.621 \\
        Lang            & 5   & 0.001 & 0.029 & 0.999 & 0.953 & 0.971 & 0.962 \\
        Mockito         & 2   & 0.000 & 0.000 & 1.000 & 0.900 & 1.000 & 0.945 \\
        \hline\hline
        Total           & 168 & 0.168 & 0.050 & 0.841 & 0.404 & 0.950 & 0.510 \\
        \hline\hline
    \end{tabular}
}
\caption{Average method accuracy of minimization on correctly executed subjects compiled using JDK 17 at source level 
17.}
\label{table:rq3-jdk17-java17-method-success}
\end{table}

One key observation from the above tables is that depending on the project, the accuracy and precision of our technique 
vary widely. From a manual investigation, we note that these subjects often contain code that relies on runtime values 
to determine branches to take and types to instantiate. Since our technique is not context-sensitive, we default to 
assuming that all runtime branches within a reachable method may be reachable, causing a higher false-positive rate in 
those subjects. This is especially true for Closure and JacksonDatabind, where the library performs parsing and 
transformation on input mainly consisting of structured text. This will be further discussed in Section 
\ref{discussion}.

Moreover, we can also see that subjects from Compress and Closure have a higher false-negative rate compared to 
subjects from other projects. We also manually investigate the cause of false negatives, and we can conclude that this 
is mainly due to two factors. Firstly, the decision of which program declarations to remove can differ even if the 
final execution result is the same, as some declarations are only retained for compilability and do not affect 
execution. Secondly, there are cases where a no-argument constructor with an empty body is determined to be unreachable 
and is removed, but at runtime the constructor is invoked via Java Reflection. Since the behavior of the original 
constructor is the same as the default constructor generated by Java, the test case will continue executing with no
changes to the behavior of execution.

While the technique appears to perform better overall in a Java 8 environment compared to a Java 17 environment, this 
is only due to the different subjects used for Java 8 and Java 17, as evident by the same result obtained for the same 
subjects in Codec.

\subsection{RQ4: Technique Overhead}

\paragraph{Experiment Setup} To evaluate the additional overhead required by our technique, we take all subjects of RQ2 
and time how long it takes for the technique to process each snapshot. We also take the time required to compile each 
snapshot to provide context to the overhead of our technique.

\begin{table}
\centering
\resizebox{\textwidth}{!}{
    \begin{tabular}{ || c  c | c | c  c || }
        \hline\hline
        Project & Total Count & Compilation & Baseline & Member-Granular \\
        \hline\hline
        Codec       & 20  & 1.5 & 0.374 (25\%)  & 0.497 (33\%)    \\
        Collections & 4   & 3.7 & 2.050 (55\%)  & 10.646 (288\%)  \\
        Lang        & 67  & 3.4 & 1.701 (50\%)  & 2.196 (77a\%)    \\
        Math        & 39  & 1.1 & 2.436 (221\%) & 12.049 (1095\%) \\
        \hline\hline
        Total       & 130 & 2.6 & 1.728 (66\%)  & 5.151 (198\%)   \\
        \hline\hline
    \end{tabular}
}
\caption{Time taken in seconds for minimization on subjects compiled using Java 8, source level 1.7.}
\label{table:exec-time-jdk8-java7}
\end{table}

\begin{table}
\centering
\resizebox{\textwidth}{!}{
    \begin{tabular}{ || c  c | c | c  c || }
        \hline\hline
        Project & Total Count & Compilation & Baseline & Member-Granular \\
        \hline\hline
        Closure         & 511 & 5.2 & 15.395 (296\%) & 20.379 (392\%) \\
        Codec           & 20  & 1.7 & 0.344 (20\%)   & 0.405 (24\%)   \\
        Collections     & 7   & 2.1 & 2.078 (99\%)   & 8.583 (409\%)  \\
        Compress        & 103 & 1.3 & 1.126 (87\%)   & 1.703 (131\%)  \\
        JacksonDatabind & 219 & 2.2 & 13.863 (630\%) & 16.697 (759\%) \\
        JacksonXml      & 12  & 2.4 & 0.983 (41\%)   & 0.676 (28\%)   \\
        Lang            & 12  & 0.9 & 1.831 (203\%)  & 1.854 (206\%)  \\
        Mockito         & 67  & 5.5 & 0.923 (17\%)   & 1.283 (23\%)   \\
        \hline\hline
        Total           & 951 & 3.9 & 11.710 (300\%) & 15.174 (389\%) \\
        \hline\hline
    \end{tabular}
}
\caption{Time taken in seconds for minimization on subjects compiled using Java 17, source level 17.}
\label{table:exec-time-jdk17-java17}
\end{table}

\paragraph{Results} the results are shown in Tables {\ref{table:exec-time-jdk8-java7}} and 
{\ref{table:exec-time-jdk17-java17}}. 

We can observe that the execution time of each snapshot differs for different projects. This is because the majority of 
time is spent doing two things: determining the set of reachable methods by virtual dispatch (occurs in the mark phase) 
and tracing the reachability graph to determine if the program declaration should be removed (occurs in the sweep 
phase).

This also explains why the baseline is sometimes slower than our technique despite its simplicity. The baseline 
technique needs to scan through the entire class when finding reachable declarations, meaning that even if only a 
method is reachable in the entire class, the cost of scanning for reachable declarations is equal to if all methods are 
reachable. This is in contrast with our member-granular technique, where only reachable declarations will have their 
body scanned for other reachable declarations.

\section{Discussion}
\label{discussion}

\subsection{Applications and Implications}

As demonstrated by RQ2 and RQ3, test dependency minimization is highly effective in repairing broken snapshots due to 
changes in Java versions, restoring compilability in all cases, and reproducing the expected test result in over 80\% 
cases. Moreover, the results of RQ1 also suggest that the issue of broken snapshots is likely to grow worse as newer 
versions of Java are released and older versions are deprecated. Although \mod{our demonstration is only able to show 
the effectiveness of our technique in some scenarios}, we nevertheless believe that test dependency minimization can be 
used with other build repair tools (such as BuildMedic \cite{macho2018automatically} and LibCatch 
\cite{zhong2021migrating}) and other automated compilation error repair tools to maximize the effectiveness of 
automatic snapshot compilation and bug bisection. \mod{We also believe that future works may address the problem of 
automatically applying patches from newer to older revisions for automated bisection.}

Another use case where test dependency minimization can benefit is the collection of bug corpora. Research in the field 
of software engineering often collects software repositories from online sources such as GitHub to be used for 
evaluating novel techniques. The authors for Defects4J noted that the usage of real bugs in software projects is often 
preferred over synthetic bugs \cite{just2014defects4j}. Moreover, \cite{just2014defects4j} and \cite{saha2018bugs} both 
state that as part of the bug collection pipeline, the authors will remove all snapshots that cannot be compiled from 
the subject pool, as the dataset is aimed to support the evaluation of dynamic analysis techniques. However, this also 
leads to lowered diversity for the collected dataset since the uncompilable bugs are excluded from the dataset. We 
believe that by using a workflow similar to the one proposed for automated bug bisection, more snapshots can be made 
available for researchers, allowing for novel works to be more comprehensively evaluated.

Finally, we believe that test dependency minimization can be useful for refactoring operations. As software systems grow
larger over time, there has been an increased focus on keeping software projects maintainable and comprehensible 
\cite{baqais2020automatic}. While many integrated development environments (IDEs) already provide some level of support 
for refactoring, we believe that test dependency minimization can be extended to supplement existing refactoring tools 
to more accurately extract necessary declarations for a given functionality in a software project.

One thing to note is that some control over the extent of minimization may be necessary for test dependency minimization
to be used in some novel bytecode-based techniques. This is because tools in some domains, especially in automated 
generation of repairs and test cases, rely on the full context of the program to discover code patterns that can be 
used 
in the generation process \cite{durieux2019empirical}.

However, while the current technique is demonstrated to be highly precise, the precision of the technique can vary 
depending on the project and each snapshot, which can be explained by the unsoundness when analyzing snapshots using 
the 
Java Reflection APIs and the lack of context sensitivity.

Regarding the issue of Java Reflection APIs, programs can use these APIs to arbitrarily load classes, invoke 
constructors and methods, and retrieve or modify the value of class fields. Since the inputs to these APIs often are 
strings, usually of identifier names, they cannot be easily inferred by static analysis techniques, which is a known 
limitation of such techniques on Java programs. While some recent works tried to address the unsoundness problem 
\cite{smaragdakis2015more}, it is unclear whether these works are beneficial to our technique, as our technique also 
requires a low false-positive rate to minimize the chance of including a program declaration that introduces 
compilation errors. One possible direction for solving this issue is to detect when a test case fails due to unsound 
analysis, and automatically add the supposedly-unreachable methods and/or constructors as an explicit entrypoint, so 
that subsequent minimization passes will unconditionally keep these declarations.

As for the issue of context sensitivity, the implementation is explicitly chosen to be context insensitive due to the 
possibility of state explosion as described in Section {\ref{directly:static_analysis}}. When determining the possible 
values of a variable, the possible values in the initializer and all assignment expressions must be considered, meaning 
that these values need to be traced across the entire program. Moreover, arguments to method calls may be derived from 
a 
return value from the same method, meaning that the possible values need to be iteratively solved until a tight bound 
can be reached, causing the time cost to exponentially increase with program complexity. Future works may investigate 
techniques to infer the possible range of values of a variable or method return value, such as by symbolic execution.

\subsection{Threats to Validity}

\paragraph{Internal Validity} The major threat to internal validity is that the coverage-based technique is used as the 
baseline for comparison against our proposed technique. If the coverage-based technique is not implemented correctly, 
this will affect the statistical results in the evaluation. We mitigate this threat by designing test cases to verify 
that the coverage-based technique is correctly minimized, as well as performing manual comparisons between 
coverage-based and our proposed techniques if false negatives are detected.

\paragraph{External Validity} One major threat to external validity is the use of Defects4J as the dataset for 
evaluation, which may not generalize to other projects. We mitigate this threat by using all available bugs in the 
Defects4J to maximize the diversity of bugs used for evaluation.

Another major threat to external validity is that subjects in Defects4J are patched to execute under a Java 7 compiler, 
meaning that the results may be biased towards the Defects4J version of bugs rather than the original snapshots. 
However, we believe that this does not affect the evaluation of our technique, as the Java 7 compiler that was bundled 
with Defects4J supports the source level of all Defects4J subjects. Moreover, using pre-patched versions of Defects4J 
subjects can be seen as a best-case scenario since these bugs have already been verified to work under Java 7, meaning 
that any compilation errors that arise from using a newer version of Java will occur regardless.

\paragraph{Construct Validity} A major threat to construct validity is that only test results are used to verify the 
correctness of the program behavior. This may be insufficient as a correct execution of a test case does not imply the 
execution trace of the original and minimized snapshots are the same, as methods may be invoked in a different order or 
the number of iterations executed for a loop may differ. We partly mitigate this threat by running each project 
revision 
on all triggering tests, as well as running tests on both the buggy and fixed versions of the snapshot. To fully 
mitigate this threat, Cobertura can be run over the original and minimized snapshots to obtain a Hit Count Vector, and 
the hit count for each statement can be compared. Note that although this cannot detect the order in which methods are 
invoked, this is regardless more rigorous than only using the result of assertions.

\section{Related Works}
\label{related-works}

\paragraph{Reasons for Snapshot Breakage} An empirical study conducted by Tufano \etal of 100 Java projects shows that 
out of around 220,000 snapshots, only 38.13\% of all project snapshots can be compiled, with less than 36\% of the 75\% 
of oldest revisions being compilable \cite{tufano2017there}. This study also mentioned that 14\% of snapshots are 
uncompilable because of parsing or compilation errors, and this result is replicated by Hassan \etal, where 9 and 3 
projects out of 91 build failures from the latest project snapshots are due to incorrect Java Development Kit (JDK) 
version and compilation errors respectively \cite{hassan2017automatic}. However, it is important to note that both 
studies were conducted in 2017 when around 95\% of software projects still use Java 7 and 8 in their implementation 
\cite{javaversion2017}; Since then, Java 11 LTS is the most popular targeted Java version \cite{javaversion2022}.

\paragraph{Repairing Build Breakage} Research on repairing build breakage is often motivated by Continuous Integration 
(CI) failures. Zhang \etal \cite{zhang2019large} performed an empirical study investigating the reasons and 
corresponding fixes for compilation errors in CI, where its categorization of compilation errors is useful for 
categorizing compilation errors in broken snapshots. While the reasons for CI build failures may intersect with reasons 
for snapshot build failures, CI environments are often similar to the expected compile-time and runtime environment of 
the project, as the main goal of CI is to catch unexpected errors or regressions under the execution environment of the 
project.

Vassallo \etal outlined a technique to aid the debugging of build failures by summarizing the error message and 
providing hints to the developer for possible solutions \cite{vassallo2018break}, but is not useful in an automatic 
bisection context because these hints are only useful for a developer to manually investigate and make changes. Macho 
\etal developed a technique to automatically fix dependency-related build breakage in Maven 
\cite{macho2018automatically}, which may be used in conjunction with our technique to fix a majority of snapshot build 
failures. 

\paragraph{Bug Corpus Collection} One of the most used bug corpus for Java is Defects4J \cite{just2014defects4j}, due 
to its ease of use and reproducible nature of bugs. During the bug collection process, the authors mentioned several 
conditions for a bug to be included in its dataset, of relevance is that the bug is ``reproducible using the project's 
build system and an up-to-date JVM''. This indicates that during the collection of subjects in the Defects4J dataset, 
uncompilable revisions are eliminated automatically, which may exclude bugs present in older snapshots that use old 
versions of Java, and thus hinder the diversity of bugs included in the dataset.

\paragraph{Automatic Program Repair} Automatic program repair has been investigated by many previous works primarily to 
fix bugs in source code. As summarized in \cite{goues2019automated}, the three main repair techniques utilize either 
heuristics, constraints, or learning. However, automatic program repair aims to address the logical correctness of a 
program rather than the compilation correctness of the program, therefore this work targets a different category of 
issues.

On the other hand, there have also been works investigating the automatic repair of build scripts, such as 
\cite{hassan2018hirebuild}, which outlines using historic versions of build scripts to perform build repair. However, 
changes in the Java version used in compiling the project is an extrinsic change and does not involve the modification 
of the build script, so it is unclear whether historic versions of build scripts will contain the correct fix to repair 
such a broken snapshot.

\paragraph{Partial Program Analysis} Partial program analysis refers to the analysis of code snippets that may 
constitute a part of a bigger program, but cannot be compiled standalone. GRAPA \cite{zhong2017boosting} is a recent 
work that addresses this problem by inferring and creating the missing fragments of the code snippets, such that tools 
designed for complete programs can also be used on partial programs. While both GRAPA and our technique aim to address 
the problems caused by uncompilable code, GRAPA focuses on the addition of code to resolve missing names in snippets, 
whereas our technique focuses on the removal of code to resolve generic compilation errors, as we observe that removal 
of unresolved or ambiguous names are effective ways to addressing compilation issues.

\paragraph{Compilation Error Repair} Automatic repair of compilation errors is a commonly investigated topic as it 
occurs frequently during the software development process. Current state-of-the-art techniques utilize machine learning 
techniques to mine for correlations between compilation error messages and their respective fixes 
\cite{chhatbar2020macer, mesbah2019deepdelta}. While these techniques show promising results for identifying potential 
fixes with little code change, it is unclear whether the proposed fixes can still retain the execution trace of the 
program, making the effectiveness of these fixes dubious in the context of automated bug bisection. Our work addresses 
this limitation by making the preservation of program behavior an explicit goal and proposing reachability rules to 
ensure that the dependent components of all entrypoints are not modified during the repair process.

\mod{Another line of research specifically addressing compilation errors caused by API-breaking changes is LibCatch
\cite{zhong2021migrating}, which defines a set of migration operators for addressing different types of compilation 
errors. While the approach is demonstrated to successfully address real migrations, some migration actions may cause 
behavioral changes to client code, such as the generation of method stubs. As such, our technique can be used to 
complement LibCatch either by acting as an initial minimizer for reducing the sources of compilation errors before 
applying LibCatch, or to provide additional reachability information to decide whether the compilation error can be 
addressed by code removal rather than migration.}

\paragraph{Class Dependency Analysis} Class dependency analysis has also been investigated by many previous works 
primarily to improve developers' understanding of a software system, where there have been works as early as 2002 which 
investigate the use of class dependency information for visualization purposes \cite{barowski2002extraction}. 
Regardless, a common feature of all dependency extraction techniques is requiring that the bytecode must be present, as 
most works either directly analyze bytecode for dependencies due to the straightforward but consistent format 
\cite{barowski2002extraction} and the potential lack of source code for some software 
\cite{nanthaamornphong2015bytecode}, or utilizes the Java Debug Interface at runtime to obtain dependency information 
\cite{pinzger2008tool} since it contains the most accurate representation of class state and dependencies at any given 
point of the program execution. However, since this work aims to repair uncompilable snapshots, there is no bytecode 
information that can be used for analysis, and therefore existing techniques cannot be used for class dependency 
analysis.

Our work addresses this limitation by implementing class dependency analysis at source-level rather than 
bytecode-level, allowing class dependency information to be available to partial and uncompilable programs.

\paragraph{Software Minimization} Minimization of software projects has been investigated extensively in previous 
works. One line of research in minimization utilizes delta debugging (\textit{ddmin}), including C-Reduce 
\cite{regehr2012test}, Peres \cite{sun2018perses}, and Chisel \cite{heo2018effective}. These works perform reduction by 
successively removing chunks of a test case until the test case is minimal while still exhibiting the desired behavior. 
However, when applied to test dependency minimization on uncompilable revisions, since the desirable behavior (i.e. 
whether a test case would pass or fail under a revision) is not known, \textit{ddmin} cannot be used.

Another technique used for minimization is to reuse existing compiler optimizations to inline function calls and 
subsequently perform dead-code elimination to remove unneeded components of the program, and is used in the field of 
code debloating. Works in this line of research include Trimmer \cite{sharif2018trimmer} and Occam
\cite{malecha2015automated}. The key difference between this line of work and our work is that compiler optimization 
techniques are often implemented after the compiler frontend completes parsing, semantic analysis, and IR lowering, the 
input program must be compilable to take advantage of these optimizations. Similarly, Razor \cite{qian2019razor} adopts 
an approach where debloating is performed on bytecode to enable debloating at runtime; However, this has the same 
drawback as compilation techniques, where successful compilation of the input program is required.

A more recent work related to minimization for software debloating is DomGad \cite{xin2020subdomain}, which identifies 
all likely paths taken by a subdomain of inputs, and then performs stochastic optimizations to minimize the program 
while preserving its generality. Compared to our technique, DomGad is path-sensitive and therefore may be able to 
achieve better minimization for imperative programs, whereas our technique instead focuses on devirtualization as the 
primary means of minimization. Furthermore, DomGad requires coverage data to obtain the path executed by each sample 
execution.

J-Reduce \cite{kalhauge2019binary} performs minimization of compiled classes by utilizing \textit{ddmin} on the binary 
classes of a compiled program rather than individual statements within a test case, procedurally reducing the set of 
classes until a minimal set of dependent classes are found. However, J-Reduce also operates on bytecode and thus 
requires successful compilation for its use.

Our technique addresses the above issues by only using source-level techniques to perform minimization, such that it 
can be performed on uncompilable programs. Moreover, we also inject assertions to warn of undesirable behavior that may 
result from imprecisions inherent to our technique.

CodeEx \cite{zhong2023empirical} is a tool that extracts usages of APIs in projects as examples for developers using 
said APIs and its implementation comprises of \textit{removers} that resolves any compilation errors by removing 
redundant constructs surrounding the usage of the API. However, CodeEx's technique involves removing program components 
regardless of whether the component influences program behavior for a given entrypoint. Our technique addresses this by 
making the preservation of program behavior an explicit aim, similar to as described above.

\section{Conclusion and Future Work}
\label{conclusion}

Build breakages of software snapshots due to updates in Java versions are becoming more frequent due to the shortened 
release cycle and amount of changes in each major release of Java, which causes significant manual effort to be needed 
when performing bug bisection for a bug that spans multiple Java versions. 

In this paper, we analyzed bug snapshots from Defects4J to investigate the reasons breakages occur when recompiling a 
Java project under a newer version of Java. Based on this insight, we propose a novel dependency minimization technique 
to remove sources of compilation errors by removing classes and methods that are not used by any execution of 
triggering test cases in a snapshot. To our knowledge, the technique is the first work that performs 
behavior-preserving source-level minimization to address compilation errors.

Using Defects4J as our evaluation dataset, we discover that build failures after a Java compiler upgrade occur in 
12\%-47\% of Defects4J bugs, and can be separated into 4 categories: Changes to the Java Language, changes to the Java 
standard library, unsupported encoding, and unsupported build tools. Build failures are also more common when upgrading 
to the latest versions of JDK compared to an older version of JDK.

We then show that test dependency minimization can repair all broken snapshots for compilation and up to 84\% of broken 
snapshots for test execution across different JDK and source level versions, and on average achieve over 95\% method 
recall for all broken snapshots. At the same time, test dependency minimization on average takes between 0.5 to 20 
additional seconds per snapshot depending on the complexity of the project, showing that the minimization process 
introduces minimal overhead when included in a bisection process.

For future work, there are several aspects that our technique can still improve upon. Firstly, the current algorithm 
does not perform any context-sensitive analysis, meaning that methods that contain a large number of branches may 
introduce significant false positives due to the assumption that any branch in the method may be executed. Future works 
may investigate utilizing techniques such as symbolic execution to eliminate unreachable branches. 

Secondly, the current algorithm does not handle classes and methods which are instantiated or invoked using Java 
Reflection APIs, causing false negatives and unexpected execution errors; This may be solved by leveraging existing 
works on statically solving declarations used in Reflection APIs. 

Finally, since the current technique primarily uses removal to perform minimization, this technique may not be optimal 
when using the repaired snapshot for generation-based tasks such as automatic test case generation, as these tasks 
often take advantage of unused classes and methods to generate objects used as operands for the target method. 
Therefore, future work may investigate alternative implementations that balance automatically fixing compilation 
errors and retaining the maximal slice of the snapshot.

With minor improvements to the technique, we believe that test dependency minimization can also be applied to fixing 
uncompilable subjects during bug corpus collection, as well as integrated into IDEs to provide more accurate code 
extraction capabilities.

\section*{Acknowledgement}

This research is partially supported by the Hong Kong RGC/GRF grant number 16207120.



\newpage
\bibliographystyle{elsarticle-num} 
\bibliography{reference}

\end{document}